
\input phyzzx.tex
\def\sqr#1#2{{\vcenter{\vbox{\hrule height.#2pt
          \hbox{\vrule width.#2pt height#1pt \kern#1pt
          \vrule width.#2pt}
           \hrule height.#2pt}}}}
\def\square{\mathchoice\sqr66\sqr66\sqr66\sqr66}

\titlepage

\hskip 4.5in CPTH-S380.1095

\hskip 4.5in LPTENS-95-46

\hskip 4.5in Crete-95-18

\hskip 4.5in hep-th/9510112

\title{ONE-LOOP EFFECTIVE ACTION
              AROUND DE-SITTER SPACE$^*$}
\foot{
Supported in part by the EEC contracts SC1-CT92-0792,
CHRX-CT93-0340 and CHRX-CT94-0621 and by the Greek Ministry
of Research grant $\Pi ENE\Delta \;91E\Delta 358$
}

\author   { I. Antoniadis }
\address{Centre de Physique Th\'eorique, Ecole Polytechnique,$^\dagger$
 \break
91128 Palaiseau; France} \foot{ Laboratoire Propre du CNRS UPR A.0014 }


\author {J. Iliopoulos }
\address{Laboratoire de Physique Th\'eorique de l'Ecole Normale Sup\'erieure,
$^\ddagger$
 \break
24 rue Lhomond, 75005 Paris; France}
\foot{Unit\'e Propre du CNRS associ\'ee \`a l'ENS et \`a l'Universit\'e
de Paris Sud. }
\centerline {and}
\author { T.N. Tomaras$^{\S}$}  \foot{
email: tomaras@physics.uch.gr
}
 \address{Physics Department, University of Crete and \break
 Research Center of Crete, P.O. Box 2208, 710 03 Heraklion, Crete; Greece }
 \vskip 1.5cm
\centerline{\bf abstract}

The non-local one-loop contribution to the gravitational effective action
around de Sitter space is computed using the background field method with
pure trace external gravitational fields and it is shown to vanish.
The calculation is performed in a generic covariant gauge and the result is
verified to be gauge invariant.

\endpage
          \chapter{Introduction}

\REF\rMyr{N. P. Myrhvold, {\it Phys. Rev.} {\bf D28} (1983) 2439 }

\REF\rF{L. H. Ford, {\it Phys. Rev.} {\bf D31} (1985) 710}

\REF\rM{E. Mottola, {\it Phys. Rev.} {\bf D31} (1985) 754; ibid {\bf D33}
(1986) 1616 and 2136; P.O. Mazur and E. Mottola, {\it Nucl. Phys.} {\bf B278}
(1986) 694}

\REF\rAM{I. Antoniadis and E. Mottola, {\it Journal Math. Phys.} {\bf 32}
(1991)
 1037 }

\REF\rAIT{I. Antoniadis, J. Iliopoulos and T. N. Tomaras,
{\it Phys. Rev. Lett.} {\bf 56} (1986) 1319}

\REF\rTW{N. Tsamis and R. Woodard, {\it Commun. Math. Phys.} {\bf 162} (1994)
 217; {\it Class. Quantum Grav.} {\bf 11} (1994) 2969}

\REF\rAlT{B. Allen and G. Turyn, {\it Nucl. Phys.} {\bf B292} (1987) 813}

\REF\rFIT{E. G. Floratos, J. Iliopoulos and T. N. Tomaras, {\it Phys. Lett.}
{\bf 197B} (1987) 373}

\REF\rR{B. Ratra, {\it Phys. Rev. } {\bf D31} (1985) 1931 }

\REF\rdWitt{B. S. DeWitt, {\it Dynamical Theory of Groups and Fields},
Gordon and Breach, New York, 1965; G. 't Hooft and M. Veltman,
{\it Ann. Inst. Henri Poincare} {\bf 20} (1974) 69  }

\REF\rAb{L. F. Abbott, {\it Nucl. Phys.} {\bf B185} (1981) 189 }

\REF\rBoul{D. F. Boulware, {\it Phys. Rev. } {\bf D23} (1981) 389 }

\REF\rAlc{B. Allen, {\it Phys. Rev. } {\bf D34} (1986) 3670}

\REF\rDP{E. D'Hoker and D. H. Phong, {\it Rev. Mod. Phys.} {\bf 60}
(1988) 917}

\REF\rP{J. Polchinski, {\it Phys. Lett.} {\bf B219} (1989) 251}

 \par
It has been shown that quantum gravity in de Sitter space-time
suffers from several pathological features
[{\rMyr} - \rTW].
In particular, the graviton propagator grows at large distances
[\rAM, \rAlT, \rFIT].
Although this behavior points towards an instability of de Sitter space,
the precise significance of this result is still unclear.
An immediate consequence is that even tree-level scattering amplitudes
diverge [\rFIT]. However, the physical meaning of a quantity such as a
scattering amplitude is obscure in a space with a horizon. Therefore we have
looked for another quantity in which the large distance behavior of the
propagator could manifest itself and which is independent of the definition of
asymptotic states. As such we have chosen the one-loop effective action
suitably
defined to be gauge invariant.

In order to illustrate our purpose, let us
consider first the simpler problem of a massless scalar field coupled to
an external de Sitter gravitational background:
$$
{\cal L}=-\sqrt{-g} \{ {1\over 2}(D\Phi)^2 + {\zeta\over 12}{\cal R}\Phi^2\}
\eqn\lscalar
$$
where $\zeta = 0$ corresponds to the massless minimal coupling
and $\zeta = 1$ to the
conformal one. The first case is expected to have similar large-distance
properties with fully quantized gravity [\rF].
Strictly speaking, the propagator of a
massless minimally coupled scalar field around a de Sitter background is
singular because of the zero mode of the Laplacian operator on the
four-sphere. However, dropping the zero mode, one obtains a two-point function
which grows logarithmically at large distances [\rR] in the same way as the
transverse and traceless part of the graviton propagator.  Therefore one
could think
that this case represents a simple model for quantum gravity.
In the conformally invariant coupling on the other hand,
the propagator vanishes at large distances
[\rR] and no pathological behavior is expected.
We have computed in both cases the
effective action for the classical gravitational field. We write
$$ g_{\mu \nu} = g^{dS}_{\mu \nu} + H_{\mu \nu}  $$
where $g^{dS}_{\mu \nu}$ is the de Sitter metric.

The effective action $\Gamma$ is a functional of $g_{\mu \nu}$.
The first non-trivial
term in its expansion in powers of $H_{\mu \nu}$ is the second-order
one which has in an
obvious notation the
form:
$$
\Gamma^{(2)}[H] = \int H^{\mu \nu}(x)\;
\gamma^{(2)}_{\mu \nu, \mu^{\prime} \nu^{\prime}}(x, x^{\prime})\;
 H^{\mu^{\prime} \nu^{\prime}}(x^{\prime})\;
dV(x) \;dV(x^{\prime})
\eqn\etwothree
$$

\noindent
where $dV(x)$ is the de Sitter volume element at $x$ and indices are contracted
with the de Sitter metric. We wish to compute $\gamma^{(2)}$
for $x \ne x^{\prime}$ and in particular to determine its
behavior for large separations of $x$ and $x^{\prime}$. The computation is
straightforward but, for simplicity we give here the result for $H_{\mu
\nu}(x)$ restricted to the special form
$$
H_{\mu \nu}(x) = {1\over 4}\; g^{dS}_{\mu \nu}(x)\; H(x)
\eqn\etwofive
$$
The result is
$$
\gamma^{(2)}=2(\zeta-1)^2[(G_{;\mu;\nu'})^2 +4\zeta (G_{;\mu})^2 +4\zeta^2G^2]
+{\rm local\ \ terms}
\eqn\rscalar
$$
where $;$ denotes covariant differentiation with respect to the de Sitter
background and $G(x,x^{\prime})$ is the scalar propagator $\Delta_{m^2}$
defined in eq.(A.23) with $m^2=-2\zeta$ in units such that the radius of the
Euclidean four-sphere equals one.

As expected, for the conformally invariant coupling $\zeta=1$, $\gamma^{(2)}$
is local and is given by the conformal anomaly. For the minimal
coupling $\zeta=0$, $G$, defined by the sum over all non-zero modes on $S^4$,
is given by
$$
G={1\over (4\pi)^2}\ [\ {1\over(1-z)}-2\ln (1-z)\ ]
\eqn\scalarprop
$$
The invariant distance $z(x,x')$ is obtained by analytic continuation from the
geodesic distance $\mu(x,x')$ on the four-sphere by $z=\cos^2(\mu/2)$. Note
that $z=1$ corresponds to $x=x'$, while the large distance limit is given by
$z\rightarrow\infty$. $\gamma^{(2)}$ is not local any more and in fact
using $\scalarprop\;$ and $\rscalar\;$ one obtains up to local terms:
$$
\gamma^{(2)}={6\over (4\pi)^4}\ \ [\ {1\over (1-z)^4}+{1\over (1-z)^3}+{1\over
(1-z)^2}\ ]
\eqn\bescalar
$$
At large $z$ the leading term is $z^{-2}$ which can be interpreted as a
logarithmically divergent correction to the effective cosmological constant
in the
one-loop effective action. The $z^{-3}$ and $z^{-4}$ terms represent similarly
the logarithmically divergent corrections to the $\cal R$ and ${\cal R}^2 $
terms.
We want to examine whether a similar behavior is
present in the case of quantum gravity.

The plan of the paper is as follows: In the next section we define the
effective action in the context of the background field method we will
use throughout.
Section 3 contains the actual computation of the quadratic term in the
one-loop approximation for the special case of pure trace background
gravitational field. The gauge invariance of the result has been explicitly
verified.
An important technical result is the expression of the graviton propagator
in an arbitrary covariant gauge and is presented in the Appendix.
We chose to work with the space-time signature ($-$ + + +).

             \chapter{ The model }

\par

We are interested in studying quantum gravity around de Sitter space.
So, we consider the lagrangian density:
$$
{\cal L}=-\sqrt{-g} \;( {\cal R }- 2 \Lambda)
\eqn\etwoone
$$

\noindent
with $\Lambda$ the cosmological constant.
It is known [{\rdWitt} - \rBoul] that in the
background field method the
effective-action
is automatically invariant under the so-called background gauge
transformations and that it becomes
invariant also with respect to quantum gauge transformations if one
chooses the background field to be a solution of the classical
equations of motion [\rBoul].

We thus expand the metric as
$$
g_{\mu \nu} = g^{dS}_{\mu \nu} + H_{\mu \nu} + h_{\mu \nu}
\eqn\etwotwo
$$

\noindent
where
$g^{dS}_{\mu \nu}+H_{\mu \nu} \equiv {\overline g}_{\mu \nu}$ is the full
gravitational background and $h_{\mu \nu}$ the quantum fluctuating
field.
$H_{\mu \nu}(x)$ is restricted to be a solution of the classical equations
of motion
$$
{\overline {\cal R}}_{\mu \nu} - {1\over 2}{\overline g}_{\mu \nu}
{\overline{\cal R}}
 + \Lambda{\overline g}_{\mu \nu}= 0
\eqn\etwofour
$$

These equations show that in quantum gravity the requirement of gauge
invariance relates the possible terms in the effective action as a
function of the background field. In other words, the computation of
$\Gamma^{(2)}$ will not allow us to identify the behavior of each
individual term. This problem becomes more severe if one considers,
for simplicity, pure trace background fields as in eq. $\etwofive$. At
the linearized level the equations of motion $\etwofour$ yield:
$$
H_{;\mu ;\nu} = - \;g^{dS}_{\mu \nu} \; H
\eqn\etwosix
$$

These equations are too restrictive. The only solution is a
superposition of
the five conformal Killing fields of de Sitter space.
With this form of $H$ the effective action becomes just a function of
five real variables and is found to be local.
We conclude that the pure trace part of $\Gamma^{(2)}$ for the theory
$\etwoone$
cannot teach us anything about the large distance properties of
the theory.

A possible way  to overcome this difficulty is to
consider the case of quantum gravity coupled to matter fields which add
in the right hand side of eq. $\etwofour$ the matter energy momentum
tensor. In this note we shall consider the simple problem of gravity
coupled to a massive scalar field $\Phi$ which does not change the
large distance properties of the theory.
We thus consider the model
$$
{\cal L}\;=\; -\sqrt{-g}\; [(1+ a \Phi) {\cal R} - 2 \Lambda (1+2 a \Phi)
- {1\over 2} ((D \Phi)\sp2 - \mu\sp2 \Phi\sp2)]
\eqn\etwoseven
$$
Notice that the term linear in $\Phi$ is multiplied by the trace of the
field equation $\etwofour$. This guarantees that the equations of
motion derived from $\cal L$ above still admit the solution
$[ g_{\mu\nu}(x)=g^{dS}_{\mu\nu}(x), \Phi=0 ]$. We are going to study
perturbatively the fluctuations around it.
As for the graviton field, we also decompose
the scalar field $\Phi$ into a background $\phi$ and a fluctuating part
$\varphi$.

The effective action is now a functional of $H_{\mu\nu}$ and $\phi$.
The computation of the full $\Gamma^{(2)}$
is rather lengthy, so, as a first step, we shall again consider
the simpler case
for which $H_{\mu \nu}(x)$ is a pure trace $H$.
$H$ and $\phi$ satisfy the linearized equations of motion
$$
H(x) = - 4 \;a \;\phi(x)
\eqn\etwoeight
$$
$$
(\;\square + {{\mu\sp2 - 12 a\sp2}\over {1 - 3 a\sp2}})\; \phi(x) = 0
\eqn\etwonine
$$
\noindent
which admit in de Sitter space a complete set of wave solutions.
\vskip 1cm

          \chapter {The effective action}

          \section{ The Feynman rules}

We now proceed with the computation of the effective action of the model
$\etwoseven$. The quadratic part of $\etwoseven\;$
in the quantum fields, diagonalized by the field redefinition
$$
h_{\mu \nu}  \to h_{\mu \nu} - a \varphi {\overline g}_{\mu \nu}
\eqn\ehnew
$$

\noindent
takes the form ( in our units $\Lambda = 3$ ):
$$\eqalign{
{\cal L}^Q = - \sqrt{-g} \;\; [&\;{1\over 4} ({\bar D}_{\mu} h)\sp2 +
{1\over 2} ({\bar D}_{\mu} h^{\mu \nu})\sp2 -
{1\over 4} ({\bar D}_\lambda h_{\mu \nu})\sp2 -
{1\over 2} ({\bar D}_\mu h) ({\bar D}_\nu h^{\mu \nu}) +
{3\over 2} h_{\mu \nu}^2 -
{3\over 4} h\sp2 \cr
&+ {1\over 8} {\bar{\cal R}} (h^2 - 2 h_{\mu \nu}^2) +
{1\over 2} {\bar{\cal R}}_{\mu \nu} (h^\mu_\alpha h^{\alpha\nu} - h h^{\mu
\nu})
-{1\over 2} {\bar{\cal R}}_{\mu \nu\lambda\rho} h^{\mu\rho} h^{\nu\lambda} \cr
&- {1\over 2} (1 + 3 a\sp2) ({\bar D}_\mu {\varphi})\sp2 - {1\over 2} (\mu\sp2
-12 a\sp2) {\varphi}\sp2 \;] \cr
}
\eqn\eLQ
$$

\noindent
In $\eLQ$ the covariant derivatives ${\bar D}$, the Riemann tensor
and the contractions
of indices are computed with the full
background gravitational field ${\overline g}_{\mu \nu}$.
$h$ is the trace of $h_{\mu \nu}$.

In the Appendix we compute the graviton and the ghost propagators
$G_{\mu\nu\mu'\nu'}(z)$ and $Q_{\mu\mu'}(z)$, respectively, in a
general class of gauges defined in terms of the redefined graviton
field $\ehnew$ by
$$
{\cal L}_{GF} = - {1\over {2 \alpha}} \;\sqrt{-\bar g}\;
({\bar D}^\nu h_{\mu \nu} - \xi \; {\bar D}_\mu h)\sp2
\eqn\eLGF
$$
To leading order in the quantum fields,
the redefined graviton transforms under an infinitesimal general coordinate
transformation with parameters $\epsilon_\mu(x)$ according to
$$
\delta h_{\mu \nu} = {\bar D}_\mu \epsilon_\nu +
{\bar D}_\nu \epsilon_\mu +
a {\bar g}_{\mu \nu} \epsilon^\lambda {\bar D}_\lambda \phi
$$
and this in turn leads to the corresponding quadratic ghost Lagrangian:
$$
{\cal L}_{\rm ghost} = - \sqrt{-\bar g}
({\bar D}^\mu {\bar \eta}^\nu) \bigl[{\bar D}_\mu {\eta}_\nu
+ {\bar D}_\nu {\eta}_\mu -
2 \xi {\bar g}_{\mu\nu} {\bar D}_\lambda \eta^\lambda
+a(1-4\xi) {\bar g}_{\mu\nu} \eta^\lambda {\bar D}_\lambda \phi \bigr]
\eqn\eLghost
$$

The result for the graviton propagator can be cast in the form:
$$
G=G^{(0)}+\alpha G^{(1)}
\eqn\gravprop
$$
with $G^{(0)}$ given in eqs. (A.10), (A.15), (A.16) and $G^{(1)}$ in (A.19).
The calculations simplify if one uses the value $\xi=1/4$. As shown in
the Appendix, this is one of the special
values for which the propagator on $S^4$ has a zero-mode.
However, a similar zero-mode
appears in the ghost propagator $Q$ in eq. (A.24). We will explicitly
show below that they cancel in
the one-loop calculation [\rAlc]. Notice, that this problem of zero modes is
unavoidable for any value of $\xi$ for, even when the graviton equation
has no zero modes, that of the ghost always has. In the
intermediate steps we shall use, as propagators, the inverse of the
corresponding wave operators in the non-zero-mode sub-spaces.

The results for $G$ and $Q$ are:
$$
G^{(0)}=G^{(0)}_{TT}+G^{(0)}_{PT} \eqn\eGo
$$
with the transverse traceless part $G^{(0)}_{TT}$ given in eq. (A.15) and
the pure trace part $G^{(0)}_{PT}$ given by:
$$
G^{(0)}_{PT} = g^{dS}_{\mu \nu} g^{dS}_{\mu^\prime \nu^\prime}
\sum_{n\ne 1} {{\varphi}_n {{\varphi}^\prime}_n  \over
{6 (\lambda^{(0)}_n + 4)}} +
\beta_1 \sum_{i=1}^5 {\chi_{1i}}_{\mu\nu}
{\chi^\prime_{1i}}_{\mu^\prime \nu^\prime}
\eqn\eGoPT
$$
The part containing the longitudinal pieces $G^{(1)}$ and $Q$ take the form:
$$
G^{(1)}= \sum_{n\geq1} {2\over{\lambda^{(1)}_n+3}} V_{n\mu\nu} \;
V^\prime_{n\mu^\prime\nu^\prime} + {16\over 9}\sum_{n\geq2}
{{\varphi_{n;\mu;\nu} \varphi_{n;\mu^\prime;\nu^\prime}}\over{\lambda^{(0)}_n
(\lambda^{(0)}_n+4)^2}}
\eqn\eGi
$$
$$
Q = \sum_{n\geq1} {{\xi_{n\mu} \xi_{n\mu^\prime}} \over
{\lambda^{(1)}_n + 3 }} - {2\over 3} \sum_{n\geq 2} {{\varphi_{n;\mu}
{\varphi^\prime}_{n;\mu^\prime}} \over {\lambda^{(0)}_n
(\lambda^{(0)}_n + 4)}}
+ \beta_2 \sum_{j=1}^{15} K_{j\mu} K^\prime_{j\mu'}
\eqn\eQ
$$
where $\beta_{1,2}$ are arbitrary parameters, which
should not appear in any physical quantity. This is a further check of
gauge invariance.
The $\chi_{1i}^{\mu\nu}$ are the five zero-modes of the graviton kinetic
operator ${\cal D}_{inv}+(1/\alpha){\cal D}_{GF}$ in the gauge $\xi=1/4$,
and they are given in the Appendix.
$\chi_{1i}^{\mu\nu} = (1/2) g^{\mu\nu} \phi_{1i} $ with
$\partial_\mu \phi_{1i}/2$ the five proper conformal Killing fields
of de Sitter space.
$K_{j\mu}$ denote collectively its fifteen conformal Killing fields:
the ten Killing vectors $\xi^i_{0\mu}$ (i=1,2,...10), together with the
above five proper conformal Killing vectors. They are all zero-modes of the
ghost kinetic operator in the $\xi=1/4$ gauge.
Note that because of the zero-modes equations (A.5) and (A.22) get modified to:
$$\eqalign{
{\cal D}_{inv} G^{(0)} + {\cal D}_{GF} G^{(1)} &=
{\bf 1} - \sum_{i=1}^5 \chi_{1i} \chi_{1i}^\prime  \cr
{\cal D}_{GF} G^{(0)} = {\cal D}_{inv} G^{(1)} &= 0  \cr
{\cal D}_{gh}Q &= {\bf 1} - \sum_{j=1}^{15} K_j K_j^{\prime}
}
\eqn\eProp
$$

We wish to study the large distance behavior of the
non-local part of the effective action $\gamma^{(2)}$.
So, it will not
be necessary to consider diagrams with $\varphi$ internal lines, since
these are suppressed,
and furthermore, for the evaluation of
the graviton contribution we shall
only need the three-point vertices $H-h-h$ and $\phi-h-h$.

The $H-h-h$ vertex comes from ${\cal L}^E =
-\sqrt{-g} ( {\cal R} - 2 \Lambda) $
as well as from the gauge-fixing term. We find:
$$
{\cal V}^E_{Hhh} = -
{1\over4}\; H \;\bigl[\;{1\over4} h^{\mu\beta;\alpha}
h_{\mu\beta;\alpha} - {1\over2} h^{\mu\alpha;\beta} h_{\alpha\beta;\mu} +
{1\over2} h_{;\alpha} {h^{\beta\alpha}}_{;\beta} -
{1\over4} h_{;\alpha} h^{;\alpha} -
  (h^{\alpha\mu} {h_{\mu\beta}}^{;\beta})_{;\alpha}
$$
$$
+ {1\over2} (h^{\alpha\beta} {h_{\alpha\beta}}^{;\mu})_{;\mu} +
{1\over2} (h^{\alpha\beta} h_{;\beta})_{;\alpha} \;\bigr]
\eqn\ethreefour
$$
$$ \eqalign{
{\cal V}^{GF}_{Hhh} &= {1\over{2 \alpha}}H \;\bigl[\;{3\over4}
\;({h^T_{\alpha\beta}}^{;\beta} )^2 + {1\over2} h^{T\beta}_\alpha\;
{h^{T\alpha\mu}}_{;\mu;\beta} \;\bigr] \cr
&= {1\over{2 \alpha}}H \;\bigl[\;{3\over4}
\;({h^T_{\alpha\beta}}^{;\beta} )^2 + {1\over2} h^{\alpha\beta}\;
({\cal D}_{GF} h)_{\alpha\beta} \bigr]
\ \equiv\ {1\over\alpha}{\cal V}^{(1)}
}
\eqn\eVi
$$
\noindent
where $h^T_{\mu\nu} \equiv h_{\mu\nu} - (1/4) \;g^{dS}_{\mu\nu}\; h$
denotes the traceless part of $h_{\mu\nu}$ and the operator ${\cal D}_{GF}$
is given in eq. (A.4). The expressions for the various
pieces of the vertex are by definition up to the factor $\sqrt{-g_{dS}}$
the corresponding terms in the expansion of the lagrangian density.

Similarly we obtain the $\phi-h-h$ vertex:
$$
{\cal V}_{\phi hh} = - \;a \;\phi \;\bigl[ {1\over4} (h_{;\mu})^2
+ {1\over2} ({h^{\alpha\beta}}_{;\alpha})^2 - {1\over4} (h^{\alpha\beta;\mu})^2
- {1\over2} h_{;\beta} {h^{\beta\alpha}}_{;\alpha} -
{1\over2} (h^{\alpha}_{\beta;\mu} h^{\beta\mu})_{;\alpha}
$$
$$
- {3\over2} (h^\alpha_{\beta;\alpha} h^{\beta\mu})_{;\mu} +
(h^{\alpha\beta} h_{;\alpha})_{;\beta}
- {1\over2} (h h_{;\alpha})^{;\alpha} +
(h^{\alpha\beta} h_{\alpha\beta;\mu})^{;\mu}
$$
$$
+ {1\over2} (h {h^\alpha_\mu}^{;\mu})_{;\alpha} +
h_{\alpha\beta}^2 - h^2 \bigr]
\eqn\ethreesix
$$

The vertices $\ethreefour$ and $\ethreesix$ are combined using the
equation of motion of the background fields $\etwoeight$. The result is:
$$ \eqalign{
{\cal V}^{(0)} &= - {1\over 4}H[ -{1\over 2}
h^{\alpha\beta}( \square -2)h_{\alpha\beta} + {1\over 2}h( \square +1)h
+h^{\beta\nu}h^\alpha_{\beta ;\alpha ;\nu} -
{1\over 2}h^{\alpha\beta}h_{;\alpha ;\beta} -
{1\over 2}h {h^{\alpha\beta}}_{;\alpha ;\beta}] \cr
&= - {1\over 4}H h^{\alpha\beta} ( {\cal D}_{inv} h )_{\alpha\beta}
}
\eqn\eVo
$$
where ${\cal D}_{inv}$ is the wave operator given in eq. (A.4)
and corresponds by definition to the invariant part of the
action (A.1).

Finally, the $H-ghost-antighost$ vertex is obtained from the expansion
of eq. $\eLghost$ and is given by:
$$
{\cal V}^{(G)} = \; H\; \bigl[ -{1\over2}
\bar\eta^{\alpha;\beta} (\eta_{\alpha;\beta} + \eta_{\beta;\alpha}) +
{1\over4} {\bar\eta^\alpha}_{\;;\alpha} {\eta^\beta}_{;\beta} \bigr]
\eqn\eVG
$$

A comment about notation is in order: Having expanded
to linear order with respect to the background fluctuations $H$ and $\phi$
and to second order in the quantum fields,
the expressions for
the vertices contain only contractions and covariant differentiations
with respect to the de Sitter background. Thus, in the rest of the
present paper we will simplify our notation and use
$g_{\mu\nu}$ for the de Sitter metric and both $;$ or $D_{\mu}$,
depending on notational convenience,
for the covariant derivative with respect to it.

 \section{ The $\alpha-$independence}

Using the propagators and the vertices above we calculate the
one-loop effective action. It is easier to organize the calculation
as an expansion in powers of the gauge parameter $\alpha$. In this way
the gauge invariance of the result will be manifest.

The first term is proportional to $\alpha^{-2}$. It is generated by the
diagram having two ${\cal V}^{(1)}$ vertices given in eq. $\eVi$,
and two $G^{(0)}$ propagators.  This term vanishes because
${\cal V}^{(1)}$  is proportional to ${h^{T\mu\nu}}_{;\nu}$ and the traceless
part of $G^{(0)}$ has no longitudinal piece.

The second term is proportional to $\alpha^{-1}$. It is the sum of two
diagrams, one with two ${\cal V}^{(1)}$ vertices, one $G^{(0)}$ and one
$G^{(1)}$ propagators and another one with one ${\cal V}^{(1)}$ and one
${\cal V}^{(0)}$ vertices and two $G^{(0)}$ propagators. The second
can be shown to vanish with an argument similar to the one given
above. After a straightforward calculation, the first one is found to
be proportional to
$$
G^{(0)\mu\nu\mu'\nu'}_{TT}
{{}_T{G^{(1)}_{T\mu\rho\mu'\rho'}}^{;\rho;\rho'}}_{;\nu;\nu'}
$$
where ${}_TG^{(1)}_T$ denotes the traceless part of $G^{(1)}$ with respect
to both pairs of indices. Using the expression $\eGi$ one can show that
${}_TG^{(1)}_T$ satisfies the following identity:
$$
{}_T{G^{(1)}_{T\mu\rho\mu'\rho'}}^{;\rho;\rho'}=-{\bf 1}_{\mu\mu'}
+\sum_{j=1}^{15}K_{j\mu}K^\prime_{j\mu'}
\eqn\eGiid
$$
where ${\bf 1}_{\mu\mu'}$ is the $\delta$ function for vectors in
de Sitter space. Taking two more derivatives with
respect to $\nu$, $\nu'$, symmetrizing in the two pairs of indices,
subtracting the traces and using the defining properties of the
conformal Killing fields, one can easily show that this diagram also
vanishes, up to possible local terms.

In order to finish with the explicit demonstration
of the $\alpha-$invariance of our result, we consider the terms
with positive powers of $\alpha$
in the effective action. First, the one with $\alpha^2$ arises
entirely from the
graph with two
${\cal V}^{(0)}$ vertices $\eVo$ and two $G^{(1)}$ propagators.
Its contribution is proportional to ${\cal D}_{inv} G^{(1)}$
and it is zero due to eq. $\eProp$.

Finally, two graphs contribute to the ${\cal O}(\alpha)$
terms in the effective action. Firstly, the one with
one vertex ${\cal V}^{(0)}$,
one vertex ${\cal V}^{(1)}$ and two propagators $G^{(1)}$. Like in
the previous graph the vertex ${\cal V}^{(0)}$ with two
longitudinal propagators attached to it vanishes identically.
Secondly, the graph with two vertices
${\cal V}^{(0)}$, one propagator $G^{(1)}$ and one $G^{(0)}$.
Ignoring the contractions which vanish by the previous argument
we find that its contribution is proportional to
$$ G^{(1)} {\cal D}_{inv}^\prime {\cal D}_{inv} G^{(0)} $$
Using eq. $\eProp$ and the relation
${\cal D}_{inv} \chi_{1i} = 0$, it is straightforward to
show that the result vanishes up to local terms.

This concludes our explicit demonstration of the $\alpha-$
independence of the ${\cal O} (H^2)$ part of the one-loop
effective action we are computing.
The use of the equations of motion (2.6) was crucial
for the above result [\rBoul].
Had we allowed for generic $H, \phi$ backgrounds we would have
obtained an $\alpha-$dependent answer.

\section{ The ${\cal O} (\alpha^0) $ term }

Finally, as we will show below the remaining ${\cal O} (\alpha^0) $
term of the
non-local part of the one-loop effective action density $\gamma^{(2)}$
is zero. Since we are interested here only in the non-local piece
of $\gamma^{(2)}$, we shall systematically ignore the local
contributions in all intermediate steps.

Four graphs contribute to the above quantity:

1) The graph with both vertices ${\cal V}^{(0)}$ and two
graviton propagators $G^{(0)}$.

We use the vertex ${\cal V}^{(0)}$ $\eVo$ and perform
the contractions to obtain:
$$
{\gamma^{(2)}}_{V_0 V_0} = {1\over{16}} \bigl[ G^{(0)}
{\cal D}_{inv}^\prime {\cal D}_{inv} G^{(0)} + ({\cal D}_{inv} G^{(0)})
({\cal D}_{inv}^\prime G^{(0)}) \bigr]
$$
Using $\eProp$ and ${\cal D}_{inv} \chi_{1i} = 0$
one can show that the first term vanishes and
${\gamma^{(2)}}_{V_0 V_0}$ simplifies to
$$
{\gamma^{(2)}}_{V_0 V_0} = {1\over{16}} ({\cal D}_{inv} G^{(0)})
({\cal D}_{inv}^\prime G^{(0)})
\eqn\eVoVoa
$$

With the $G^{(0)}$ and ${\cal D}_{inv}$ given in $\eGo$,
(A.15), $\eGoPT$ and (A.4) we find
$$
{\cal D}_{inv} G^{(0)} = \sum_{n\ge0} {h^{TT}_{n\mu\nu}
h^{\prime TT}_{n\mu^\prime\nu^\prime}} - {1\over3}
\bigl(D_\mu D_\nu - g_{\mu\nu} (\square + 3) \bigr) g_{\mu^\prime\nu^\prime}
\sum_{n\ne1} {{\varphi_n \varphi^\prime_n}\over{\lambda^{(0)}_n + 4} }
\eqn\eDinvGo
$$
and consequently:
$$
{\gamma^{(2)}}_{V_0 V_0} = {1\over{16}} \bigl(\sum_{r=1}^5 {\varphi_{1r}
\varphi^{\prime}_{1r}} \bigr)^2 +
{1\over{16}} \bigl(\sum_{n\ge0} {h^{TT}_n h^{\prime TT}_n} \bigr)^2
\eqn\eVoVo
$$

2) The graph with one vertex ${\cal V}^{(0)}$, the other ${\cal V}^{(1)}$
and $G^{(0)}$ and $G^{(1)}$ for the two graviton propagators.

The vertex ${\cal V}^{(1)}$ is given in $\eVi$. It contains two terms but
upon contraction
with ${\cal V}^{(0)}$ given in $\eVo$ the first term vanishes
because it contains divergences of the transverse and traceless $G^{(0)}_{TT}$.
The second simplified by the use of ${\cal D}_{inv} G^{(1)}=0$
leads to:
$$
{\gamma^{(2)}}_{V_0 V_1} = - {1\over{8}} ({\cal D}_{inv}^\prime G^{(0)})
({\cal D}_{GF} G^{(1)})
\eqn\eVoVia
$$
which combined with $\eProp$ and $\eDinvGo$ gives:
$$\eqalign{
{\gamma^{(2)}}_{V_0 V_1} &= {1\over{8}} \bigl[({\cal D}_{inv}^\prime G^{(0)})
({\cal D}_{inv} G^{(0)}) - \bigl(\sum_{r=1}^5 {\varphi_{1r}
\varphi^\prime_{1r}} \bigr)^2 \bigr] \cr
&= {1\over{8}} \bigl(\sum_{n\ge0} {h^{TT}_n h^{\prime TT}_n} \bigr)^2 \cr
}
\eqn\eVoVi
$$

3) The graph with both vertices ${\cal V}^{(1)}$ and two graviton
propagators $G^{(1)}$.

We carry out the contractions and we obtain
$$\eqalign{
{\gamma^{(2)}}_{V_1 V_1} = &{9\over{32}} \bigl(
{{}_TG^{(1)}_{T\mu\rho\mu'\rho'}}^{;\rho;\rho'}\bigr)^2
+ {1\over 16} G^{(1)} {\cal D}_{GF}^\prime {\cal D}_{GF} G^{(1)} +
{1\over 16} ({\cal D}_{GF}^\prime G^{(1)}) ({\cal D}_{GF} G^{(1)}) \cr
&+ {3\over 8} {G^{(1)}_{T\mu\rho\mu'\rho'}}^{;\rho'}
({\cal D}_{GF} {G^{(1)}_T)^{\mu\rho\mu'\alpha'}}_{;\alpha'}
}
\eqn\eViVia
$$
where $G^{(1)\mu\rho\mu'\rho'}_T \equiv G^{(1)\mu\rho\mu'\rho'}_T
-(1/4) g^{\mu'\rho'} g_{\alpha'\beta'}G^{(1)\mu\rho\alpha'\beta'} $
denotes the traceless part of $G^{(1)}$ with respect to the primed
indices. We then use $\eProp$ and (A.4) to show that the second and
fourth term of ${\gamma^{(2)}}_{V_1 V_1}$ vanish, while by
$\eProp$ and $\eGiid$ the remaining terms lead to
$$ \eqalign{
{\gamma^{(2)}}_{V_1 V_1} &= {1\over 16} \bigl[({\cal D}_{inv}^\prime G^{(0)})
({\cal D}_{inv} G^{(0)}) - \bigl(\sum_{r=1}^5 {\varphi_{1r}
\varphi^{\prime}_{1r}} \bigr)^2 \bigr] +
{9\over 32} \bigl(\sum_{j=1}^{15}K_{j\mu}K^\prime_{j\mu'}\bigr)^2 \cr
&= {1\over 16} \bigl(\sum_{n\ge0} {h^{TT}_n h^{\prime TT}_n} \bigr)^2 +
{9\over 32} \bigl(\sum_{j=1}^{15}K_{j\mu}K^\prime_{j\mu'}
\bigr)^2 \cr
}
\eqn\eViVi
$$

Adding the results $\eVoVo$, $\eVoVi$ and $\eViVi$ of the
three diagrams above, we obtain the total
contribution of the graviton loop:
$$
\gamma^{(2)}_{\rm graviton} = {1\over 4} \bigl(\sum_{n\ge0}
{h^{TT}_n h^{\prime TT}_n} \bigr)^2 +
{1\over 16} \bigl(\sum_{r=1}^5 \varphi_{1r} {\varphi'}_{1r} \bigr)^2
+ {9\over 32} \bigl(\sum_{j=1}^{15}K_{j\mu}K^\prime_{j\mu'}\bigr)^2
\eqn\eGravloop
$$

4) Finally, the ghost-loop i.e. the graph with both vertices
${\cal V}^{(G)}$ and two ghost propagators $Q$ given in eqs. $\eVG$ and
$\eQ$ respectively. Its contribution is
$$
\eqalign{
\gamma^{(2)}_{\rm ghost} &= - \bigl( D_{(\mu} D^{(\mu'} {Q_{\nu)}}^{\nu')}
\bigr)^2 - {1\over 16} \bigl(D_\mu D_{\mu'} Q^{\mu\mu'} \bigr)^2 +
{1\over 2} \bigl( D_{(\mu} D^{\mu'} Q_{\nu)\mu'} \bigr)^2 \cr
= &- \bigl[ D_{(\mu} D^{(\mu'} Q_{\nu)}^{\nu')}
 - {1\over4} g_{\mu\nu} D^\alpha D^{(\mu'} Q_\alpha^{\nu')}
 - {1\over4} g^{\mu'\nu'} D_{\alpha'} D_{(\mu} Q_{\nu)}^{\alpha'}
 + {1\over{16}} g_{\mu\nu} g^{\mu'\nu'} D^\alpha D_{\alpha'}
Q_{\alpha}^{\alpha'} \bigr]^2 \cr}
\eqn\eVGVGa
$$
It is easy to verify that:
$$
D^{(\mu} D^{(\mu'} Q^{\nu)\nu')} = -{1\over2} \sum_{n\ge1} {V_n^{\mu\nu}
V_n^{\mu'\nu'} } - {2\over3} \sum_{n\ge2} {{(D^\mu D^\nu \varphi_n)
(D^{\mu'} D^{\nu'} \varphi'_n)} \over {\lambda^{(0)}_n (\lambda^{(0)}_n + 4)}}
+{1\over 4}\beta_2 g^{\mu\nu} g^{\mu'\nu'}
\sum_{r=1}^5 \varphi_{1r} {\varphi'}_{1r}
$$
Subtracting the traces under both pairs of indices in order to form the
quantity in the second line of eq. $\eVGVGa$ and using the
completeness relation (A.14) we find for the non-local part of the ghost
contribution
$$
\gamma^{(2)}_{\rm ghost} = -{1\over 4}
\bigl( \sum_{n\ge0} h^{TT}_n h^{\prime TT}_n \bigr)^2
\eqn\eVGVG
$$
Note that the unphysical parameters $\beta_{1,2}$ have disappeared from
$\gamma^{(2)}_{\rm graviton}$ and $\gamma^{(2)}_{\rm ghost}$.

\vskip 1cm

  \section{ the zero-mode contribution }
  \par

It is crucial to realise at this point that the sum of
$\eGravloop$ and $\eVGVG$
is not the full answer. We are effectively computing
the path-integral over the metric, the scalar field and the ghosts of
the exponential of the action $\eLQ+\eLGF+\eLghost$.
For $\xi=1/4$ and $H_{\mu\nu}$ pure trace the ghost and the graviton
kinetic operators have zero-modes.
To linear order in $H$, they take the form
$$
\eta_{(j)}^\mu = (1+{1\over4} H) K_j^\mu \;\;\;\; j=1,2,...,15
\eqn\emodegh
$$
and
$$
h^{\mu\nu}_{(r)} = (1+{1\over4} H) \chi_{1r}^{\mu\nu} \;\;\;\; r=1,2,...,5
\eqn\emodegr
$$
respectively.

To treat properly these zero-modes special attention is required [\rDP].
Namely, one performs the path-integration over the subspace orthogonal
to the zero-modes but multiplies by an appropriate power of
the determinant of the matrix of inner products of the zero-modes.
The inner products, fixed to be ultralocal, are defined by:
$$\eqalign{
{\cal M}^{gh}_{jk} &\equiv
\langle\eta_{(j)}\vert\eta_{(k)}\rangle = \int d^4x {\sqrt{-\bar g}}
{\bar g}^{\mu\nu} \eta_{(j)\mu}\eta_{(k)\nu} \cr
{\cal M}^{gr}_{rs} &\equiv
\langle h_{(r)}\vert h_{(s)}\rangle = \int d^4x {\sqrt{-\bar g}}
{\bar g}^{\mu\lambda}{\bar g}^{\nu\rho} h_{(r)\mu\nu} h_{(s)\lambda\rho}\cr}
\eqn\einnerprod
$$
where ${\bar g}_{\mu\nu}=g_{\mu\nu}^{dS} (1+{1\over 4}H)$ denotes the
full background metric.
The functional integral is then multiplied by the factor
$$
det^{-1}{\cal M}^{gh} det^{-{1\over 2}}{\cal M}^{gr}
$$
where the minus sign in the second factor can be traced to the fact that
the action of Euclidean gravity is unbounded from below [\rP].
We expand this factor in powers of $H$ using eqs
$\emodegh$ and $\emodegr$ and we find that the contribution to
the non-local part of $\gamma^{(2)}$ cancels precisely the last two
terms of eq.$\eGravloop$.

Adding all contributions, we obtain
our result that the non-local part of $\gamma^{(2)}$ is zero!
Thus, we conclude that the pathological large-distance behavior
of the graviton propagator on the background de Sitter space does
not manifest itself in the Weyl non-invariant quadratic part of the effective
action in the one loop approximation.

What is the significance of this result? It may be an accident of the order
in which we are working and it may disappear at higher orders.
On the other hand,
it may be an indication that quantum gravity around
a de Sitter background behaves effectively like a conformally invariant
theory. In such a case the local terms should be related to the
conformal anomaly. In a pure Einstein theory, if one wants to compute
the various terms in the anomalous equation for the trace of the
energy-momentum tensor, one faces the problem of isolating the individual
contributions which was mentioned in section 2. Our method of coupling
the theory to a scalar field may provide a solution.

\vskip 1cm

      \appendix

In this Appendix we compute the graviton propagator for the class of
gauges given by $\eLGF$, as well as the propagator of the
corresponding Faddeev-Popov ghosts. We shall work in the Euclidean continuation
of de Sitter space which is $S^4$ and whose metric will be denoted by
$g_{\mu \nu}$ throughout this section.
The quadratic part of the lagrangian including the gauge-fixing is
$$ \eqalign{
{\cal L}^Q = -\sqrt{-g} \;\; [&{1\over 4} (h_{;\mu})\sp2 +
{1\over 2} ({h^{\mu \nu}}_{;\nu})\sp2 -
{1\over 4} ({h_{\mu \nu}}_{;\lambda})\sp2 -
{1\over 2} h_{;\mu} {h^{\mu \nu}}_{;\nu} - {1\over 2} (h_{\mu \nu})\sp2 -
{1\over 4} h\sp2 \cr
&+ {1\over {2 \alpha}} ({h_{\mu \nu}}^{;\nu} - \xi
h_{;\mu})\sp2 ] }
\eqn\eAone
$$

We expand the propagator $G_{\mu \nu \mu^\prime \nu^\prime}(x, x^\prime)$
in a power series of $\alpha$ and write (omitting the indices)
$$
G = G^{(0)} + \alpha G^{(1)} + \alpha\sp2 G^{(2)} + ....
\eqn\eAtwo
$$
$G$ satisfies the equation:
$$
({\cal D}_{inv}+ {1\over \alpha} {\cal D}_{GF})
(G^{(0)} + \alpha G^{(1)} + \alpha\sp2 G^{(2)} + ....) = {\bf 1}
\eqn\eAthree
$$

\noindent
where ${\cal D}_{inv}$ and ${\cal D}_{GF}$ denote the differential operators
of the gauge invariant and gauge fixing parts of the quadratic lagrangian
$\eAone\;$  respectively:
$$\eqalign{
({\cal D}_{inv} h)_{\mu\nu} &= -{1\over2} (\square - 2) h_{\mu\nu} +
{1\over2} (\square + 1) g_{\mu\nu} h + D_{(\mu} D^\alpha h_{\nu)\alpha} -
{1\over2} g_{\mu\nu} D^\alpha D^\beta h_{\alpha\beta} -
{1\over2} D_\mu D_\nu h  \cr
({\cal D}_{GF} h)_{\mu\nu} &= D_{(\mu} D^\alpha h_{\nu)\alpha} -
\xi D_\mu D_\nu h - \xi g_{\mu\nu} (D^\alpha D^\beta h_{\alpha\beta} -
\xi \square h ) }
\eqn\eAwave
$$
As usual two indices inside parentheses are symmetrized, i.e.
$T^{(\alpha\beta)} \equiv  (T^{\alpha\beta} + T^{\beta\alpha})/2 $.

Order by order in $\alpha$ equation (A.3) gives:
$$\eqalign{ \alpha^{-1}&: \hskip 1.5cm {\cal D}_{GF} G^{(0)} = 0 \cr
\alpha^0&: \hskip 1.5cm {\cal D}_{inv} G^{(0)} + {\cal D}_{GF} G^{(1)} =
{\bf 1} \cr
\alpha &: \hskip 1.5cm {\cal D}_{inv} G^{(1)} + {\cal D}_{GF} G^{(2)} = 0 \cr
\alpha^{2}&: \hskip 1.5cm {\cal D}_{inv} G^{(2)} + {\cal D}_{GF} G^{(3)} =0 \cr
&{} \hskip 3cm . \cr
&{} \hskip 3cm . \cr}
\eqn\eAfour $$

By definition $G^{(0)}$ is the graviton propagator calculated for $\alpha=0$
i.e. by imposing on the graviton quantum field the Landau-type gauge condition
$$
{h^{\mu \nu}}_{;\nu} = \xi\; h^{;\mu}
\eqn\eAfive
$$
Its computation is done independently as follows:
start with the graviton decomposition
$$
h_{\mu \nu} = {h^{TT}}_{\mu \nu} + {A^T}_{(\mu ;\nu)}
+ B_{;\mu;\nu} + {1\over4} g_{\mu \nu}  (- \square B + h )
\eqn\eAsix
$$

\noindent
and impose the gauge condition to obtain:
$$ {A^T}_\mu = 0, \hskip 0.4cm 3(\square + 4)B + (1-4 \xi) h = 0
  \eqn\eAseven $$

\noindent
Substitute $\eAsix \;$ together with $\eAseven\;$ into the
quadratic action to rewrite it
in terms of the independent fields ${h^{TT}}_{\mu \nu}$ and $B$ as:
$$
{\cal L}^Q = -\sqrt{-g} \;[\;{1\over2} {h^{TT}}_{\mu \nu} (- \square+2)
 {h^{TT}}^{\mu \nu} +{3\over{(4 \xi-1)\sp2}} B ((1-\xi)\;\square+3)\sp2
(\square
+4) B \;]
\eqn\eAeight
$$
Accordingly, the graviton propagator is decomposed as
$$
{G^{(0)}} = G^{(0)}_{TT} + G^{(0)}_{PT}
\eqn\AG
$$

It is convenient to introduce at this point [\rAlT] the
scalar, transverse-vector and spin-2 (transverse traceless symmetric tensor)
eigenfunctions of the $\square$ operator
on the sphere $S^4$ of radius $\rho$. They are given by
$$
\square \; \varphi^i_n (x) = \lambda^{(0)}_n \varphi^i_n (x)
$$
$$
\square \; \xi^{i\mu}_n (x) = \lambda^{(1)}_n \xi^{i\mu}_n (x)
\eqn\eAnine
$$
$$
\square \; h^{i\mu\nu}_{TTn} (x) = \lambda^{(2)}_n h^{i\mu\nu}_{TTn} (x)
$$
The corresponding eigenvalues and their degeneracies are:
$$
\eqalign{\lambda^{(0)}_n &= - {1\over{\rho^2}} n (n + 3) \cr
\lambda^{(1)}_n &= - {1\over{\rho^2}} (n\sp2 + 5n+3) \cr
\lambda^{(2)}_n &= - {1\over{\rho^2}} (n\sp2+7n+8) \cr } \qquad
\eqalign{g^{(0)}_n &= {1\over6} (n+1) (n+2) (2 n +3) \cr
g^{(1)}_n &= {1\over2} (n+1) (n+4) (2n+5) \cr
g^{(2)}_n &= {5\over6} (n+1) (n+6) (2n+7)  \cr }
\eqn\eAten
$$

We shall also need the following tensor functions:
$$ \chi^{\mu\nu}_n = {1\over2} g^{\mu \nu} \varphi_n  \hskip 0.8cm  n=0, 1, 2,
...  $$
$$ W^{\mu\nu}_n ={{ {\varphi_n}^{;\mu;\nu}-{1\over4} g^{\mu \nu}
\square \;\varphi_n }\over \sqrt{\lambda^{(0)}_n ({3\over4} \lambda^{(0)}_n
+ 3 \rho^{-2})}}  \hskip 0.8cm  n=2,3,4,...   \eqn\eAeleven  $$
$$ V^{\mu\nu}_n = {\xi_n^{(\mu;\nu)}\over \sqrt{-{1\over2}(\lambda^{(1)}_n
+ 3 \rho^{-2})} } \hskip 0.8cm   n=1,2,3,...  $$

\noindent
which satisfy the completeness relation
$$\sum_{n=0}^\infty {h^{TT}_n \;h^{TT^\prime}_n } + \sum_{n=1}^\infty
{V_n \;V^\prime_n} + \sum_{n=2}^\infty {W_n\; W^\prime_n} + \sum_{n=0}^\infty
{\chi_n\; \chi^\prime_n} = {\bf 1}  \eqn\completeness  $$

\noindent
where primed functions or tensors with primed indices are meant to be
evaluated at the point $x^\prime$.
In terms of these, the expressions for the two parts of the graviton
propagator $G^{(0)}$ in the Landau gauge $\eAfive$ are respectively (in our
units $\Lambda = 3$ which corresponds to $\rho = 1$):
$$ G^{(0)}_{TT} = \sum_{n=0}^\infty {{-2\over{\lambda^{(2)}_n - 2}}
 \; h^{TT}_n \;h^{TT^\prime}_n }  \eqn\eAthirteen  $$

$$ G^{(0)}_{PT} =  \sum_{n=0}^\infty
{{[(4 \xi -1) {\varphi}_n^{;\mu;\nu} + g^{\mu \nu}
((1-\xi)\square+3){\varphi}_n ] [(4 \xi -1)
{\varphi}^{\prime;\mu^\prime;\nu^\prime}_n
+ g^{\mu^\prime \nu^\prime}
((1-\xi)\square^\prime+3) \varphi^\prime_n ] }\over
{6 (\lambda^{(0)}_n + 4)
((1-\xi) \lambda^{(0)}_n + 3)\sp2}}  \eqn\eAfourteen  $$

\noindent
Notice that the $\lambda^{(0)}_n + 4$ factor in the denominator
vanishes for $n=1$. However $\varphi_1$ satisfies
$\varphi_{1;\mu;\nu} = - g_{\mu\nu} \varphi_1$ and consequently
the $n=1$ term in the sum of $\eAfourteen$ is actually zero.

Having computed $G^{(0)}$ we may use $\eAfour\;$ to solve for the remaining
pieces of the graviton propagator. First of all notice that
by definition $G^{(0)}$ satisfies ${\cal D}_{GF} G^{(0)} = 0 $.

To determine $G^{(1)}$ we start with its decomposition in terms of the
functions defined in $\eAeleven\;$
$$ G^{(1)} = \sum_n (b_n V_n \;V^\prime_n + c_n W_n\; W^\prime_n +
 d_n \chi_n\; \chi^\prime_n + e_n (W_n \chi^\prime_n + W^\prime_n \chi_n))
\eqn\eAfifteen  $$

\noindent
and determine the unknown coefficients so as to satisfy equations $\eAfour\;$.
It is straightforward to verify that the solution of the latter leads to:
$$ \eqalign{ b_n &= {2\over{\lambda^{(1)}_n+3}} \hskip 0.6cm n\geq1 \cr
d_n &= {{1\over4} \lambda^{(0)}_n\over{((1-\xi)\lambda^{(0)}_n+3})^2} \cr }
\qquad
\eqalign{c_n &={{3\over4}
{\lambda^{(0)}_n+4}\over{(1-\xi)\lambda^{(0)}_n+3}}\cr
e_n &={{[{3\over4}\lambda^{(0)} (\lambda^{(0)}_n+4)]^{1\over 2}\over
{2 ((1-\xi)\lambda^{(0)}_n+3})^2 }} \cr } \eqn\eAsixteen  $$
$$ G^{(k)} = 0  \;\;\; {\rm for \;all} \;\;\; k>1 $$
Inserting the above coefficients into the expression
$\eAfifteen\;$ we obtain the
useful form of $G^{(1)}$
$$G^{(1)}= \sum_{n\geq1}\left({2\over{\lambda^{(1)}_n+3}}
V_n^{\mu\nu} \;V^{\prime\mu^\prime\nu^\prime}_n +
{{\varphi_n^{;\mu;\nu} \varphi_n^{;\mu^\prime;\nu^\prime}}\over{\lambda^{(0)}_n
((1-\xi)\lambda^{(0)}_n+3)^2}} \right) \eqn\eAseventeen  $$

The propagator $G=G^{(0)}+\alpha G^{(1)}$ is well-defined except for the set
of special
values of $\xi=(k^2+3 k-3)/k(k+3)\;, \;\; k=1,2,...$ $[\rAlc]$.

In this class of gauges, the quadratic part of the Fadeev-Popov lagrangian
$\eLghost$ is
$$
{\cal L}^Q_{ghost} = - \sqrt{-g} \;\;{\bar \eta}^{\mu;\nu}
\;\;(\eta_{\mu;\nu} +
\eta_{\nu;\mu} - 2 \xi g_{\mu\nu} {\eta^\lambda}_{;\lambda})
\eqn\eAeighteen
$$
We decompose the ghost field into transverse and longitudinal parts
according to
$$ \eta^\mu = \eta_T^\mu + \eta^{;\mu}  $$
\noindent
and write the quadratic part of the ghost lagrangian
$$
{\cal L}^Q_{ghost} = \sqrt{-g} \left[ {\bar\eta}_T^\mu \;(\square + 3)\;
\eta^{}_{T \mu} - 2\bar\eta \;\left((1-\xi) \;\square + 3 \right)
\square \;\eta \;\right]
\eqn\eAghost
$$
We notice that the differential operators of $ {\cal L}^Q_{ghost}$ have
zero modes.
In the transverse part these zero modes are given by the ten Killing vectors
of the four-sphere,
$\xi^{i\mu}_0$ of eq.$\eAnine$, and they are present for all
values of $\xi$.  The longitudinal part has zero modes only for the set
of special values of $\xi$ mentioned above for which also the graviton
propagator has zero modes.

As a result, the ghost propagator $Q_{\mu\mu^\prime}$ is defined only in the
non-zero-mode subspace and satisfies:
$$
{{\cal D}_{gh}}^{\nu}_{\mu}Q_{\nu\mu^\prime} \equiv
\{(\square +3)\delta_{\mu}^{\nu}-2[(1-\xi)\square+3]D_{\mu}D^{\nu}\}
Q_{\nu\mu^\prime} = {\bf 1}_{\mu\mu^\prime}-
\sum_{i=1}^{10}\xi_{0\mu}^i\xi^{i\prime}_{0\mu^\prime}
\eqn\eAghostid
$$
Its explicit form is:
$$
Q_{\mu\mu^\prime} = Q^T_{\mu\mu^\prime} - {1\over6} \left(\Delta_0 -
\Delta_{3\over{1-\xi}} \right)_{;\mu;\mu^\prime}
$$
\noindent
where $\Delta_{m^2}$ denotes the propagator of a minimally coupled
scalar field of mass-squared $-m^2$
$$
\Delta_{m^2} = \sum_{n=0}^\infty {{\varphi_n \varphi^\prime_n }\over
{\lambda_n^{(0)} + m^2} }
\eqn\eAdscalar
$$
The corresponding summation in $\Delta_0$ starts from $n=1$.
$Q^T_{\mu\mu^\prime}$ is written in terms of the transverse parts of the
vector eigenfunctions $\eAnine$ as
$$
Q^T_{\mu\mu^\prime} \equiv \sum_{n\geq1} {{\xi_{n\mu} \xi_{n\mu^\prime}}
\over {\lambda^{(1)}_n + 3 }}
+\beta\sum_{i=1}^{10}\xi_{0\mu}^i\xi^{i\prime}_{0\mu^\prime}
$$
where $\beta$ is an arbitrary parameter which should not appear in any
physical quantity.

Using the above expression the ghost propagator takes its final form:
$$
Q_{\mu\mu^\prime} = \sum_{n\geq1} {{\xi_{n\mu} \xi_{n\mu^\prime}} \over
{\lambda^{(1)}_n + 3 }} - {1\over2} \sum_{n\geq1} {{\varphi_{n;\mu}
{\varphi^\prime}_{n;\mu^\prime}} \over {\lambda^{(0)}_n ((1-\xi)
\lambda^{(0)}_n + 3)}}
+\beta\sum_{i=1}^{10}\xi_{0\mu}^i\xi^{i\prime}_{0\mu^\prime}
\eqn\eAnineteen
$$
It is straightforward to carry out the summations over the modes
in the various propagators and express them in terms of the invariant
distance $z(x, x')$ and its derivatives [\rAlT]. For our purposes
though the use of these explicit formulas is not
particularly illuminating.


\ack\
We would like to thank Professor F. Englert for several helpful discussions.
I.A. and J.I. acknowledge the hospitality of the Physics Department
of the University of Crete and the Research Center of Crete,
and T.N.T. acknowledges the
hospitality of the Centre de Physique Th\'eorique of the Ecole Polytechnique
and the Laboratoire de Physique Th\'eorique of the Ecole Normale Sup\'erieure.

\refout
\vfill
\eject
\end